\documentclass[conference]{IEEEtran}
\usepackage{graphicx}
\usepackage{amsmath, amssymb}
\usepackage{url}

\usepackage{tikz}
\usetikzlibrary{positioning,arrows.meta,fit}
\usetikzlibrary{calc}
\usepackage{listings}
\usepackage{xcolor}
\usepackage{cite}

\usepackage[utf8]{inputenc}   

\usepackage{dblfloatfix}

\usepackage{hyperref}
\definecolor{codegray}{rgb}{0.5,0.5,0.5}
\definecolor{codepurple}{rgb}{0.58,0,0.82}
\definecolor{backcolour}{rgb}{0.95,0.95,0.92}

\lstdefinestyle{mypython}{
    backgroundcolor=\color{backcolour},
    commentstyle=\color{codegray},
    keywordstyle=\color{blue},
    numberstyle=\tiny\color{codegray},
    stringstyle=\color{codepurple},
    basicstyle=\ttfamily\footnotesize,
    breaklines=true,
    captionpos=b,
    numbers=left,
    numbersep=5pt,
    showspaces=false,
    showstringspaces=false,
    tabsize=4,
    language=Python
}

\begin{document}

\title{QERS: Quantum Encryption Resilience Score}

\author{
    \IEEEauthorblockN{Jonatan Rassekhnia}
    \IEEEauthorblockA{
        Luleå University of Technology (LTU) \\
        Department of Computer Science, Electrical and Space Engineering \\
        Sweden \\
        Email: rasjon-0@student.ltu.se
    }
}

\maketitle

\begin{abstract}
Post-quantum cryptography (PQC) is becoming essential for securing
Computer Systems, IoT, and IIoT systems against future quantum
threats. However, the operational cost of deploying PQC on
resource-constrained devices is still poorly understood. This paper
introduces the Quantum Encryption Resilience Score (QERS), a practical
evaluation instrument designed to quantify the impact of PQC on computing, IoT, and IIoT  communication systems. QERS integrates latency, packet reliability, CPU load, energy use, signal strength, and key size into a unified normalized score. Three complementary scoring formulas are introduced: (i) Basic, for rapid
comparisons, (ii) Tuned, for environment-specific weighting, and
(iii) Fusion, which combines normalized performance and security
sub-scores into a composite indicator. An ESP32C6-Devkit testbed was used to evaluate five PQC algorithms (Kyber, Dilithium, Falcon, SPHINCS+, and NTRU) under different wireless conditions. The results show that QERS exposes performance–security trade-offs that are difficult to observe through isolated benchmarks, providing a reproducible instrument for assessing PQC readiness in resource-limited environments.
\vspace{0.2 cm}

\end{abstract}
\vspace{0.2 cm}

\begin{IEEEkeywords}
Post-Quantum Cryptography, IoT Security, Multi-Protocol Communication, Design Science, Quantum Simulation
\end{IEEEkeywords}

\section{Introduction}
The rapid expansion of Traditional Computing, Internet of Things (IoT) and Industrial IoT (IIoT) has introduced billions of connected devices that operate under strict resource constraints, while simultaneously it relying on cryptographic mechanisms for secure communication. At the same time, advances in quantum computing threaten classical public-key cryptography, since algorithms such as RSA and ECC may become breakable once sufficiently large quantum systems emerge \cite{PhysRev.47.777, Nofer2023}. As a result, the transition toward Post-Quantum Cryptography (PQC) has become a strategic and technical priority for future secure systems \cite{jaidka2020iot, Mathur2020, 9286147, YALAMURI2022834, 11059928}.
\vspace{0.1 cm}

Although several PQC algorithms that include Kyber, Dilithium, Falcon, SPHINCS+, and NTRU, have been standardized or are currently undergoing evaluation, their \textit{operational impact} on constrained Traditional Computing and  IoT/IIoT platforms remains insufficiently characterized. Existing research primarily focuses on micro-benchmarks such as computation time, memory footprint, or throughput. However, these metrics alone do not reflect how PQC could affects end-to-end system behavior, including network latency, packet reliability, CPU utilization, energy consumption, and wireless stability \cite{moody2021nist, yashika2025iot, liu2024post}. This lack of system-level visibility complicates PQC adoption decisions, particularly across diverse hardware and protocol environments \cite{wiesmaier2021pqcmigrationcryptoagility}.
\vspace{0.1 cm}

To address this gap, this paper introduces the \textbf{Quantum Encryption Resilience Score (QERS)}, a multi-metric evaluation instrument designed to quantify the resilience of Traditional Computing and IoT/IIoT communication systems operating under PQC workloads. Rather than benchmarking cryptography in isolation, QERS integrates heterogeneous metrics, including latency, packet loss, CPU load, signal strength, energy usage, and cryptographic key parameters, into a normalized composite score. The framework provides three complementary scoring modes \cite{Bhol2020CyberSecurityMetrics, Ameyed2023}:
\vspace{0.1 cm}

\begin{itemize}
    \item \textbf{Basic:} rapid comparison of PQC impact across protocols.
    \item \textbf{Tuned:} environment-aware weighting for specific deployment priorities.
    \item \textbf{Fusion:} combined performance and security sub-scores producing a unified readiness indicator.
\end{itemize}
\vspace{0.1 cm}

An ESP32-based test is employed to evaluate five PQC algorithms across different wireless conditions. Experimental results demonstrate that QERS exposes performance and security trade-offs that remain hidden when metrics are analyzed individually, offering a reproducible means of assessing PQC readiness in constrained environments.
\vspace{0.1 cm}

\subsection{Research Problem}
Despite progress in PQC standardization, there is no unified, reproducible measurement instrument capable of evaluating PQC across multiple operational dimensions on real Computing, IoT, and IIoT platforms. As a result, engineers and manufacturers lack decision support when estimating feasibility and deployment cost at scale \cite{al2020internet}.
\vspace{0.1 cm}

\subsection{Research Question}
\begin{itemize}
\item How can a unified and reproducible measurement instrument be designed to evaluate PQC algorithms across Computer Systems, IoT, and IIoT platforms while incorporating latency, memory usage, energy behavior, Key-Bytes, CPU usage, RSSI and packet-level communication characteristics?
\end{itemize}
\vspace{0.1 cm}

\subsection{Contributions}
This research introduces the \textbf{Quantum Encryption Resilience Score (QERS)}, a layered evaluation instrument designed to measure PQC impact under realistic operating constraints. The main contributions are:

\begin{enumerate}
    \item A modular scoring instrument supporting Basic, Tuned, and Fusion modes for progressively deeper analysis.
    \item A normalization and weighting methodology integrating latency, CPU usage, signal reliability, key size, packet behavior, and energy metrics into performance and security sub-scores.
    \item A practical implementation on ESP32 hardware for real-world experimentation, avoiding purely simulated environments.
    \item An empirical evaluation of five PQC algorithms across different wireless scenarios.
    \item A reproducible and extensible framework designed to support future PQC testing, comparison, migration toward quantum-resilient systems, and deployment in resource-limited environments.
\end{enumerate}
\vspace{0.1 cm}

\section{Related Work \& Background}
\subsection{Post-Quantum Cryptography}
Post-Quantum Cryptography (PQC) aims to develop cryptographic schemes that remain secure against adversaries equipped with large-scale quantum computers \cite{chen2016report}. The National Institute of Standards and Technology (NIST) is actively standardizing PQC algorithms to replace classical public-key systems that are vulnerable to Shor’s algorithm \cite{borges2020comparison}. 
\vspace{0.1 cm}

For constrained Computer Systems, IoT, and IIoT environments, the most relevant PQC families include Kyber, Dilithium, Falcon, SPHINCS+, and NTRU \cite{10048976, Barhoumi2025}. These schemes differ significantly in underlying hardness assumptions, key sizes, ciphertext or signature sizes, and computational requirements. Table~\ref{tab:pqc_algorithms} summarizes representative trade-offs between efficiency and security levels.
\vspace{0.1 cm}

\begin{table}[h!]
\centering
\caption{Selected PQC Algorithms}
\label{tab:pqc_algorithms}
\scriptsize
\begin{tabular}{|c|c|c|c|c|p{1.2cm}|}
\hline
\textbf{Alg.} & \textbf{Type} & \textbf{Hard Prob.} & \textbf{Key (B)} & \textbf{CT/Sig (B)} & \textbf{Notes} \\
\hline
Kyber      & KEM & Module-LWE   & 800–1500  & 768–1088   & Efficient, balanced \cite{Barhoumi2025} \\
\hline
Dilithium  & Sig & Module-LWE   & 1312–2544 & 2420–3500  & Large signatures \cite{Barhoumi2025} \\
\hline
Falcon     & Sig & NTRU lattice & 897–1280  & 690–1024   & Compact, float-heavy \cite{Barhoumi2025} \\
\hline
SPHINCS+   & Sig & Hash-based   & 32–64     & 8000–16000 & Stateless \cite{Barhoumi2025} \\
\hline
NTRU       & KEM & NTRU lattice & 1138–1420 & 1138–1420  & High security \cite{Barhoumi2025} \\
\hline
\end{tabular}
\end{table}
\vspace{0.1 cm}

\subsection{Constraints in Computing and IoT/IIoT Security}
Computing systems operate on deterministic states and can only approximate quantum behavior through simulation \cite{Tacchino2020QuantumSim}. They inherently lack the ability to natively represent quantum properties such as superposition and entanglement, which limits their ability to efficiently solve certain problem classes \cite{preskill2012quantumcomputingentanglementfrontier}. These architectural constraints are particularly relevant when deploying PQC on Computer Systems, IoT, and IIoT devices, which rely on classical processors with limited computational, memory, and energy resources \cite{Nagy01042007}.
\vspace{0.1 cm}

Computer Systems, IoT and IIoT devices are typically resource-constrained, often operating with tight memory budgets, limited CPU capacity, and restricted network bandwidth \cite{9139976, 9381665, 8472907}. Intermittent connectivity and packet loss further complicate secure communication. Consequently, PQC algorithms with large keys or computational overhead can increase latency, degrade throughput, and accelerate battery depletion \cite{cryptography9020032}.
\vspace{0.1 cm}

\subsection{Quantum \& PQC Benchmarks}
Early studies primarily evaluated PQC on traditional compute platforms, CPUs, GPUs, and FPGAs primarily focusing on runtime performance, memory requirements, and throughput \cite{cryptography9020032}. More recent research expands toward benchmarking methodologies, risk frameworks, and quantum-aware IoT evaluation models. Lall et al. \cite{lall2025reviewcollectionmetricsbenchmarks} and Acuaviva et al. \cite{acuaviva2025benchmarkingquantumcomputersstandard} survey performance metrics and benchmarking approaches for quantum computing platforms, while Soubra et al. \cite{metrics2020009} investigate measurement frameworks at both hardware and software layers. Quantum risk frameworks such as QARS address cryptographic exposure and migration readiness \cite{electronics14173338}.
\vspace{0.1 cm}

In the IoT domain, Rehman and Alharbi \cite{smartcities8040116} propose QESIF, integrating quantum key distribution with intrusion detection, and Hossain et al. \cite{hossain2024quantumedgecloudcomputingfuture} explore quantum-edge architectures for improved resilience. Ameyed et al. \cite{Ameyed2023} emphasize evaluation models focusing on measurable security quality indicators.
\vspace{0.1 cm}

Recent applied research demonstrates real deployment feasibility. QRoNS demonstrates PQC integration in high-performance networking, achieving near line-rate IPsec throughput using Kyber1024 and Dilithium5 on NVIDIA BlueField-3 DPUs \cite{electronics14214234}. Similarly, the QORE framework enables end-to-end PQC protection in 5G/B5G networks with minimal latency penalties \cite{rathi2025qorequantumsecure}. EPQUIC evaluates PQC within TLS and QUIC stacks on embedded platforms, revealing significant differences across algorithms regarding latency overhead and protocol behavior \cite{Dong2025EPQUIC}. Finally, the Quantum Approximate Optimization Algorithm (QAOA) is explored as a promising approach for combinatorial optimization on future quantum hardware \cite{Gupta2021}.
\vspace{0.1 cm}

\subsection{Evaluation Metrics and Decision Models}
Evaluating PQC in constrained environments requires metrics that extend beyond raw cryptographic execution time \cite{Marchsreiter2025QuantumBlockchain}. Relevant indicators include latency, jitter, throughput, packet loss, energy consumption, and memory usage \cite{iot6040062, pr9112084, 10317890}. Multi-Criteria Decision Analysis (MCDA) and Multi-Criteria Decision Making (MCDM) approaches are widely applied to rank cryptographic alternatives under competing system constraints \cite{10763879, Triantaphyllou2000, ishizaka2013multi}. Min–max normalization is frequently used to harmonize heterogeneous metrics, enabling meaningful comparisons across devices and protocols \cite{book}.
\vspace{0.1 cm}

Within MCDA, metrics can be organized into conceptual categories:

\begin{itemize}
\item \textbf{Cost Criteria:}
Derived from classical cost–benefit analysis, cost criteria capture negative operational impacts such as latency, CPU usage, energy consumption, packet loss, and cryptographic overhead \cite{Nash1975CostBenefit, iot6040062, aslam2025quantum}. Higher values indicate performance penalties that should be minimized.
\vspace{0.1 cm}

\item \textbf{Benefit Criteria / Security Gain:}
Security gain metrics reflect improvements in robustness, reduced exposure, and proven resistance to attack, informed by frameworks such as SAEM \cite{1007971, butler2003security}. These metrics also incorporate organizational perspectives on information security value \cite{Dhillon2006ValueFocusedIS}. Parameters such as proven resistance ($P_r$), key size ($K$), cryptographic robustness, and reliability directly contribute to overall security posture \cite{10401941}.
\vspace{0.1 cm}

\item \textbf{Performance Criteria:}
Grounded in general systems theory \cite{Shaw01112009}, performance criteria assess operational efficiency through metrics such as latency ($L$), CPU usage ($C$), energy consumption ($E$), packet loss ($P_{loss}$), and signal strength ($R$). These indicators support integrated evaluation of IoT/IIoT platforms across different deployments \cite{demir2025performanceanalysisindustrydeployment, app14124994, boo1, book}.
\vspace{0.1 cm}

\end{itemize}

Min–max normalization further enables consistent scoring across diverse measurement scales \cite{patro2015normalizationpreprocessingstage}, forming the basis for the Quantum Encryption Resilience Score (QERS) introduced in this study.
\vspace{0.1 cm}

\section{Methodology (QERS Formulation)}
The Quantum Encryption Resilience Score (QERS) framework provides a structured method for evaluating the operational resilience of computer systems, IoT, and IIoT platforms when post-quantum cryptography is applied. Instead of assessing cryptographic performance in isolation, QERS integrates multiple system-level metrics such as latency, packet reliability, CPU usage, energy consumption, signal strength, cryptographic overhead, and key size into a unified scoring process.
\vspace{0.1 cm}

These raw metrics are first normalized to a common scale and then combined through three progressively detailed scoring models (Basic, Tuned, and Fusion). This layered design enables researchers and practitioners to compare algorithms and deployments consistently, quantify trade-offs between performance and security, and better understand how PQC affects real-world devices and networks.
\vspace{0.1 cm}

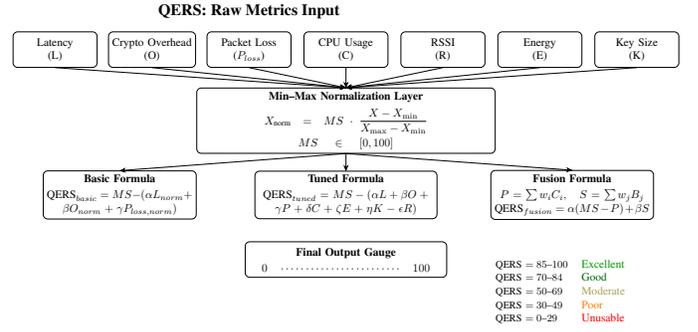
\begin{figure}[ht]
\centering
\resizebox{\columnwidth}{!}{%
\begin{tikzpicture}[
    box/.style={rectangle,draw,rounded corners,
        minimum width=2.6cm, minimum height=1.1cm, align=center},
    >=Stealth
]

\node at (6,3.2) {\Large \textbf{QERS: Raw Metrics Input}};

\node[box] (L) at (0,2) {Latency\\(L)};
\node[box] (O) at (3,2) {Crypto Overhead\\(O)};
\node[box] (P) at (6,2) {Packet Loss\\($P_{loss}$)};
\node[box] (C) at (9,2) {CPU Usage\\(C)};
\node[box] (R) at (12,2) {RSSI\\(R)};
\node[box] (E) at (15,2) {Energy\\(E)};
\node[box] (K) at (18,2) {Key Size\\(K)};

\node[box, text width=9cm, align=center] (norm) at (9,-0.2)
{\textbf{Min--Max Normalization Layer}\\[6pt]
$X_{\text{norm}} = MS \cdot
\dfrac{X - X_{\min}}{X_{\max} - X_{\min}}$ \\[4pt]
$MS \in [0,100]$
};

\foreach \n in {L,O,P,C,R,E,K}
  \draw[->] (\n.south) -- (norm.north);

\node[box, text width=4.5cm, align=center] (basic) at (2,-2.5)
{\textbf{Basic Formula}\\[2pt]
$\text{QERS}_{basic}
 = MS - (\alpha L_{norm} + \beta O_{norm} + \gamma P_{loss,norm})$
};

\node[box, text width=5.4cm, align=center] (tuned) at (9,-2.5)
{\textbf{Tuned Formula}\\[2pt]
$\text{QERS}_{tuned}
 = MS - (\alpha L + \beta O + \gamma P
 + \delta C + \zeta E + \eta K - \epsilon R)$
};

\node[box, text width=4.8cm, align=center] (fusion) at (16,-2.5)
{\textbf{Fusion Formula}\\[2pt]
$P=\sum w_i C_i,\quad S=\sum w_j B_j$\\
$\text{QERS}_{fusion}= \alpha(MS-P)+\beta S$
};

\draw[->] (norm.south) -- (basic.north);
\draw[->] (norm.south) -- (tuned.north);
\draw[->] (norm.south) -- (fusion.north);

\node[box, text width=6cm, align=center] (gauge) at (9,-4.5)
{\textbf{Final Output Gauge}\\[2pt]
$0 \quad\cdots\cdots\cdots\cdots\cdots\cdots\cdots\cdots\quad 100$
};

\node[align=left] at (15.7,-5.5) {
\begin{tabular}{@{}l l@{}}
\small QERS $=85$--$100$ & \textcolor{green!60!black}{Excellent} \\
\small QERS $=70$--$84$  & \textcolor{green!40!black}{Good} \\
\small QERS $=50$--$69$  & \textcolor{yellow!60!black}{Moderate} \\
\small QERS $=30$--$49$  & \textcolor{orange}{Poor} \\
\small QERS $=0$--$29$   & \textcolor{red}{Unusable} \\
\end{tabular}
};

\end{tikzpicture}%
}

\caption{Conceptual QERS framework showing metrics, normalization, score variants, and readiness interpretation.}
\label{fig:qers_final}
\end{figure}

\subsection{Metrics Definition: Quantum Encryption Resilience Score (QERS)}
\vspace{0.1 cm}

The \textbf{Quantum Encryption Resilience Score (QERS)} primarily evaluates communication protocols operating under Post-Quantum Cryptography (PQC) by measuring key performance metrics such as latency ($L$), packet reliability ($P_{loss}$), RSSI ($R$), PQC key bytes ($K$), and CPU usage ($C$). Metrics are assessed under multiple protocol loads and cryptographic operations, including encryption and decryption scenarios with varying key sizes derived from PQC key size recommendations. 
\vspace{0.1 cm}

The proposed QERS framework follows a Multi-Criteria Decision Analysis (MCDA) methodology and implements a Multi-Criteria Decision Making (MCDM) model for score aggregation and ranking. MCDA is used to structure the evaluation problem and define performance and security criteria, while the MCDM formulation computes a composite resilience score through weighted normalization and aggregation \cite{Bhol2020CyberSecurityMetrics, belton2002multiple, Tasopoulos2022PQTLS13
 }.
\vspace{0.1 cm}

QERS is intended to help identify potential security and performance gaps, including protocol-specific weaknesses that fall outside the immediate scope of this research. The primary objective is to provide a reproducible instrument that reflects the operational stability and efficiency of multi-protocol IoT, IIoT, and computer system devices \cite{app14104215}.
\vspace{0.1 cm}

\noindent The QERS computation employs a \textbf{Multi-Criteria Decision Model (MCDM)}, where the parameters $L, O, P_{loss}, C, R, E,$ and $K$ are weighted according to system priorities \cite{boo1, pr11051313}. Weighting allows QERS to adapt to specific scenarios such as real-time operation, energy-constrained deployments, or balanced general-purpose IoT systems. To refine these coefficients empirically, supervised learning techniques such as linear regression and random forest regression may be applied to experimental datasets or simulated environments, depending on the strategy used to observe performance behavior under varying encryption key sizes and QERS interoperability \cite{geron2019hands, chollet2021deep}.
\vspace{0.1 cm}

\noindent Each metric can be normalized using min–max scaling to ensure fair cross-protocol comparison:
\vspace{0.1 cm}

\[
X_\text{norm} = MS \times \frac{X - X_\text{min}}{X_\text{max} - X_\text{min}}
\]
\cite{boo1, book, belton2002multiple}

In this study, the min–max bounds ($X_{\min}$ and $X_{\max}$) were derived from the experimental dataset rather than theoretical limits, ensuring that normalization reflects realistic operating conditions.

\vspace{0.1 cm}

Here, $MS$ represents the chosen maximum scale (e.g., 100), replacing the fixed percentage scale. This normalization approach supports consistency across heterogeneous metrics and helps reduce bias caused by large value disparities. QERS adopts an MCDA structure for evaluating complex decisions using multiple criteria and can also be applied in group-based decision systems where several evaluators independently contribute their judgment \cite{Tawakuli2023IoT, belton2002multiple}. The \textit{Handbook on Constructing Composite Indicators and User Guide} presents the min–max normalization concept, where minimum and maximum values are determined across contexts to produce normalized indicators ranging from 0 to $MS$ \cite{book}.
\vspace{0.1 cm}

\noindent QERS can be structured hierarchically:

\begin{itemize}
\item \textbf{Basic QERS:} Evaluates core performance metrics ($L, O, P_{\text{loss}}$) using a normalized score for quick PQC readiness comparison.
\item \textbf{Tuned QERS:} Incorporates CPU usage, RSSI, energy consumption, and key size ($C, R, E, K$) with normalized scoring for more detailed analysis.
\item \textbf{Fusion QERS:} Combines normalized \textit{Performance} and \textit{Security} subscores ($P$ and $S$) to generate a comprehensive score for IoT/IIoT devices. In Fusion QERS, $MS$ represents the \textbf{maximum score} on the chosen scale (e.g., 100) and serves as the reference for normalizing performance metrics. By subtracting the normalized performance subscore $P$ from $MS$, higher QERS values indicate stronger protocol resilience while maintaining comparability across metrics and deployment environments.
\end{itemize}
\vspace{0.1 cm}

\begin{table}[h!]
\centering
\caption{Mapping of MCDM Symbols to QERS Metrics}
\label{tab:mcdm_qtrs_mapping}

\begin{tabular}{|c|p{0.68\columnwidth}|}
\hline
\textbf{Symbol} & \textbf{Description} \\ \hline
QERS & Quantum Encryption Resilience Score (0--100) \\ \hline
$X_{\min}, X_{\max}$ & Minimum/maximum values for normalization \\ \hline
$X_{\text{norm}}$ & Min--max normalized value \\ \hline
$x_{ij}$ & Criterion $j$ value for alternative $i$ \\ \hline
$r_{ij}$ & Normalized version of $x_{ij}$ \\ \hline
$w, w_i$ & Weight of criterion $i$ \\ \hline
$\sum_i w_i = 1$ & Weights for performance metrics:
$L, J, P_\text{loss}, C, E$ \\ \hline
$\sum_j w_j = 1$ & Weights for security metrics:
$K, R, P_r, Co$ \\ \hline
$A$ & Set of alternatives (protocols) \\ \hline
$X$ & Set of evaluation criteria \\ \hline
$\alpha \dots \eta$ & Decision weights in QERS \\ \hline
L & Latency \\ \hline
J & Jitter \\ \hline
$P_{\text{loss}}$ & Packet loss ratio \\ \hline
O & Cryptographic overhead \\ \hline
C & CPU utilization \\ \hline
R & RSSI / signal strength \\ \hline
E & Energy consumption \\ \hline
K & Key size \\ \hline
Co & Cryptographic overhead (memory/bandwidth/compute) \\ \hline
Pr & Proven resistance level \\ \hline
\end{tabular}

\leavevmode\vspace{4pt}\newline
\begin{minipage}{0.95\linewidth}
\footnotesize \textit{Sources:} \cite{boo1, book, belton2002multiple}
\end{minipage}

\end{table}

\noindent Weighting coefficients $(\alpha, \beta, \gamma, \delta, \epsilon, \zeta, \eta)$ allow QERS to be tailored for real-time, energy-constrained, or balanced systems \cite{boo1, book}. Following approaches in data fusion for similarity coefficients, the QERS weighting scheme can be adapted using fixed-loss or learned weights, enabling machine learning techniques to optimize the relative importance of each performance and security metric under different IoT/IIoT operational conditions \cite{OCAMPO2024113155, MLWeight, chen2009mlweighting, 6547983}.

\subsection{QERS Formulas with Normalized Metrics}

\subsubsection{Basic QERS}
\begin{equation}
\text{QERS}_{\text{basic}} = MS - (\alpha L_\text{norm} + \beta O_\text{norm} + \gamma P_{\text{loss,norm}}) \nonumber
\end{equation}

\begin{align}
\text{QERS}_{\text{tuned}} = MS &- (\alpha L_\text{norm} + \beta O_\text{norm} + \gamma P_{\text{loss,norm}} \nonumber\\
&+ \delta C_\text{norm} + \zeta E_\text{norm} + \eta K_\text{norm}) \nonumber\\
&+ \epsilon R_\text{norm} \nonumber
\end{align}

\subsubsection{Fusion QERS}

\paragraph{Performance Subscore:}
\begin{equation}
P = \sum_i w_i C_i^\text{norm}, \quad \sum_i w_i = 1 \nonumber
\end{equation}

\paragraph{Security Subscore:}
\begin{equation}
S = \sum_j w_j B_j^\text{norm}, \quad \sum_j w_j = 1 \nonumber
\end{equation}

\paragraph{Combined Fusion Score:}
\begin{equation}
\text{QERS}_{\text{fusion}} = \alpha (MS - P) + \beta S, \quad \alpha + \beta = 1 \nonumber
\end{equation}

\subsection{QERS Interpretation}

\begin{table}[h!]
\centering
\scriptsize
\caption{QERS Readiness Levels}
\begin{tabular}{|c|l|}
\hline
\textbf{QERS Range} & \textbf{Interpretation / Readiness Level} \\ \hline
85--100 & Excellent – strong performance under PQC load \\ \hline
70--84 & Good – operationally efficient, moderate crypto impact \\ \hline
50--69 & Moderate – noticeable degradation, optimization recommended \\ \hline
30--49 & Poor – significant performance loss, limited real-world applicability \\ \hline
0--29 & Unusable – protocol not viable without major redesign \\ \hline
\end{tabular}
\label{tab:qers_readiness}
\end{table}

\vspace{0.1 cm}

\subsection{QERS Evaluation: Basic, Tuned, and Fusion Metrics}

\textbf{Basic QERS Weighting Scenarios} \\[0.2em]
\begin{tabular}{|c|c|c|c|}
\hline
Scenario & $\alpha$ & $\beta$ & $\gamma$ \\ \hline
RT & 0.55 & 0.20 & 0.15 \\ 
EC & 0.25 & 0.45 & 0.20 \\ 
B  & 0.35 & 0.30 & 0.20 \\ \hline
\end{tabular} \\[0.2em]
\footnotesize{RT = Real-time, EC = Energy-constrained, B = Balanced; $\alpha$ = Latency, $\beta$ = Overhead, $\gamma$ = Packet Loss.} \\[0.8em]

\textbf{Tuned QERS Weighting Scenarios (Core + Auxiliary Metrics)} \\[0.2em]
\begin{tabular}{|c|c|c|c|c|c|c|c|}
\hline
Scenario & $\alpha$ & $\beta$ & $\gamma$ & $\delta$ & $\epsilon$ & $\zeta$ & $\eta$ \\ \hline
RT & 0.55 & 0.20 & 0.15 & 0.05 & 0.025 & 0.025 & 0.05 \\ 
EC & 0.25 & 0.45 & 0.20 & 0.025 & 0.025 & 0.05 & 0.05 \\ 
B  & 0.35 & 0.30 & 0.20 & 0.05 & 0.05 & 0.05 & 0.05 \\ \hline
\end{tabular} \\[0.2em]
{\footnotesize
$\alpha$ = Latency, $\beta$ = Overhead, $\gamma$ = Packet Loss, $\delta$ = CPU Usage, 
$\epsilon$ = RSSI, $\zeta$ = Energy, $\eta$ = Key Size.
}
\normalsize
\vspace{0.8em}

\begin{table}[htbp]
\textbf{Fusion QERS – Performance Subscore} \\[0.2em]
\begin{tabular}{|c|c|c|c|c|c|}
\hline
$L$ & $J$ & $P_{loss}$ & $E$ & $C$  \\ \hline
0.3 & 0.1 & 0.2 & 0.2 & 0.2 \\ \hline  
\end{tabular} \\[0.8em]
{\footnotesize Latency, Jitter, Packet Loss, Energy, CPU }
\normalsize

\vspace{0.1 cm}

\textbf{Fusion QERS – Security Subscore} \\[0.2em]
\begin{tabular}{|c|c|c|c|}
\hline
$K$ & $R$ & $P_r$ & $Co$ \\ \hline
0.25 & 0.35 & 0.25 & 0.15 \\ \hline
\end{tabular} \\[0.2em]
{\footnotesize
Key size ($K$), Algorithm robustness ($R$), Proven resistance ($P_r$), Cryptographic overhead ($Co$).
}
\normalsize

\label{tab:combined_qers}
\end{table}

\subsection{Machine Learning Support for Fusion QERS}
QERS also integrates a machine learning layer to assist in estimating the Fusion score under dynamic operating conditions \cite{JISC55}. A Random Forest regression model is used because it is robust to noise, handles nonlinear relationships, and performs well on small to medium datasets \cite{9457563,Langsetmo2023RandomForestRisk}. The model is trained using historical measurements containing latency, CPU usage, RSSI, packet loss, energy values, and key-size data, together with their analytically computed QERS Fusion scores \cite{Kaushik2023}. 
\vspace{0.1 cm}

Once trained, the model can predict an approximate Fusion value for new measurements, reducing sensitivity to short-term fluctuations \cite{Reis2019, 8481535}. This enables trend-based assessment rather than relying strictly on single-point measurements \cite{Masini2023MLTimeSeries}. Importantly, the machine learning component does not replace the QERS equations \cite{didona2015hybrid, 8778229}. Instead, it acts as a complementary estimator that enhances robustness when metrics are missing, unstable, or incomplete \cite{Carpenter2021MissingData, Thomas2021MLImputationReview}. 

\section{Proposed Framework}
In this research, a multi-layer cybersecurity framework is proposed to measure, evaluate, and predict the quality and efficiency of post-quantum cryptography (PQC) algorithms on embedded processors representing typical Computer Systems, IoT, and IIoT devices. The framework consists of three main layers:

\begin{itemize}
    \item \textbf{Data Collection Layer:} ESP32 and Raspberry Pi devices collect latency, CPU usage, RSSI, energy, and key-size metrics for different PQC algorithms.
    \item \textbf{Processing Layer:} Collected metrics are normalized, smoothed using exponential smoothing, and processed by the QERS models and a Random Forest fusion layer.
    \item \textbf{Visualization Layer:} Results are presented through a Flask-based dashboard using bar charts, heatmaps, and historical trends.
\end{itemize}
\vspace{0.1 cm}

{\normalsize
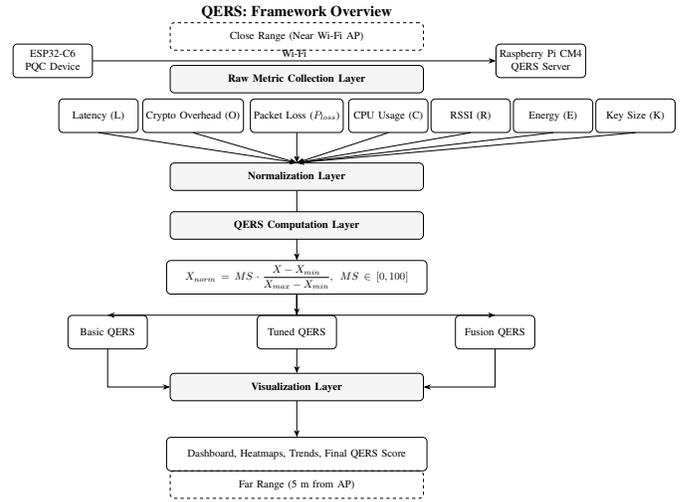
\begin{figure}[!t]
\centering

\resizebox{\columnwidth}{!}{%
\begin{tikzpicture}[
    box/.style={rectangle,draw,rounded corners,
        minimum width=2.6cm, minimum height=1.1cm, align=center},
    layer/.style={rectangle,draw,rounded corners,
        minimum width=8.3cm, minimum height=0.9cm, align=center,
        fill=gray!8},
    rangectx/.style={rectangle,draw,dashed,rounded corners,
    minimum width=8.3cm, minimum height=0.9cm, align=center},
    arrow/.style={->, thick},
    >=Stealth
]

\node at (6,4.6) {\Large \textbf{QERS: Framework Overview}};

\node[rangectx] (close) at (6,3.8) {Close Range (Near Wi-Fi AP)};
\node[rangectx] (far) at (6,-10.9) {Far Range (5 m from AP)};

\node[box] (esp) at (-2,3.0) {ESP32-C6 \\ PQC Device};
\node[box] (pi) at (14,3.0) {Raspberry Pi CM4 \\ QERS Server};

\draw[arrow] (esp.east) -- (pi.west) node[midway, above] {Wi-Fi};

\node[layer] (raw) at (6,2.4) {\textbf{Raw Metric Collection Layer}};
\node[box] (L) at (-0.5,1.2) {Latency (L)};
\node[box] (O) at (2.6,1.2) {Crypto Overhead (O)};
\node[box] (P) at (6,1.2) {Packet Loss ($P_{loss}$)};
\node[box] (C) at (9,1.2) {CPU Usage (C)};
\node[box] (R) at (11.7,1.2) {RSSI (R)};
\node[box] (E) at (14.4,1.2) {Energy (E)};
\node[box] (K) at (17.1,1.2) {Key Size (K)};

\node[layer] (normL) at (6,-0.8) {\textbf{Normalization Layer}};
\node[box, text width=8.3cm, align=center] (norm) at (6,-4.1)
{$X_{norm} = MS \cdot \dfrac{X - X_{min}}{X_{max} - X_{min}},\; MS \in [0,100]$};

\foreach \n in {L,O,P,C,R,E,K}
  \draw[arrow] (\n.south) -- (normL.north);

\draw[arrow] (normL.south) -- (norm.north);

\node[layer] (comp) at (6,-2.4) {\textbf{QERS Computation Layer}};

\node[box] (basic) at (-0.2,-5.9) {Basic QERS};
\node[box] (tuned) at (6,-5.9) {Tuned QERS};
\node[box] (fusion) at (12.5,-5.9) {Fusion QERS};

\draw[arrow] (norm.south) |- (basic.north);
\draw[arrow] (norm.south) -- (tuned.north);
\draw[arrow] (norm.south) |- (fusion.north);

\node[layer] (vis) at (6,-7.7) {\textbf{Visualization Layer}};
\node[box, text width=8.3cm, align=center] (dash) at (6,-9.9)
{Dashboard, Heatmaps, Trends, Final QERS Score};

\draw[arrow] (basic.south) |- (vis.west);
\draw[arrow] (tuned.south) -- (vis.north);
\draw[arrow] (fusion.south) |- (vis.east);
\draw[arrow] (vis.south) -- (dash.north);

\end{tikzpicture}
}

\caption{Proposed QERS framework extended with ESP32-C6 PQC device, Raspberry Pi CM4 QERS server, and close-range and far-range wireless measurement contexts.}
\label{fig:qers_framework}

\end{figure}
}
\normalsize

\vspace{0.1 cm}

To clarify how the abstract QERS framework is realized in practice, the experimental hardware and data flow are mapped onto the framework layers in Fig.~\ref{fig:qers_framework}. The ESP32-C6 executes post-quantum cryptographic algorithms and generates raw performance and wireless metrics, which are transmitted over Wi-Fi to a Raspberry Pi CM4 that performs QERS computation and visualization. Close-range and far-range wireless conditions are included to capture environmental effects on cryptographic and system behavior.

\subsection{Interpretation \& Use Cases}

The hardware setup for the proposed framework includes:

\begin{itemize}
    \item \textbf{ESP32 devices:} The ESP32C6-Devkit collect live metrics per PQC algorithm \cite{9733026}.
    \item \textbf{Raspberry Pi devices:} The Raspberry Pi CM4 aggregates data from multiple ESP32C6 units, processes metrics, and hosts the Flask dashboard \cite{simadiputra2021rasefiberry}.
    \item \textbf{Protocols tested:} Post-quantum cryptographic algorithms including \textit{Kyber, Dilithium, Falcon, SPHINCS+, and NTRU} \cite{technologies12120241, scrivano2025comparativestudyclassicalpostquantum}.
    \item \textbf{Metrics:} Key size, latency, CPU utilization, RSSI, and energy, used to compute QERS Basic, Tuned, and Fusion scores with ML-predicted fusion intervals \cite{s24092781}.
\end{itemize}

The use cases include of:

\begin{enumerate}
    \item Real-time PQC performance monitoring in IoT networks \cite{Rahmati2025, 10130471, 9570203}.
    \item Historical trend analysis for security evaluation \cite{HALDER2022351, electronics11040529}.
    \item Identification of low-performing PQC algorithms in constrained devices \cite{9328432}.
\end{enumerate}
\vspace{0.1 cm}

\subsection{Illustrative Validation / Methodology}
To validate the proposed framework, we performed representative measurements on multiple PQC algorithms using an ESP32-C6 device. Metrics were collected continuously and smoothed to reduce fluctuations \cite{gardner1985exponential,1460680}. A Random Forest model was used as a supplementary fusion layer to smooth, interpolate, and estimate QERS values from noisy or partially observed metric streams \cite{4603103,Kandhoul2021RFCSec}. Synthetic samples were generated to follow the same statistical distributions as the collected measurements in order to stabilize the learning process and prevent artificial bias.

\section{Results}
A continuous 24-hour measurement campaign was conducted using an ESP32C6-DevKit in order to illustrate the behavior of the proposed QERS framework under realistic operating conditions. 
Five post-quantum cryptographic algorithms (Kyber, Dilithium, Falcon, SPHINCS+, and NTRU) were exercised repeatedly over Wi-Fi. Each iteration recorded latency, CPU utilization, RSSI, and three QERS variants (QERS$_{Basic}$, QERS$_{Tuned}$, and QERS$_{Fusion}$), allowing observation of how QERS responds to varying system conditions.
\vspace{0.1 cm}

Two representative wireless scenarios were considered:
\vspace{0.1 cm}

\begin{itemize}
    \item \textbf{Scenario 1:} ESP32 positioned close to the Wi-Fi access point.
    \item \textbf{Scenario 2:} ESP32 positioned approximately 5 meters away.
\end{itemize}
\vspace{0.1 cm}

All measurements were recorded to CSV and processed through the QERS computation pipeline. 
All reported QERS scores were normalized to a 0–100 scale, enabling direct comparison of QERS behavior across different conditions.
\vspace{0.1 cm}

\subsection{Heatmap Analysis}
\begin{figure}[htbp]
    \centering
    \includegraphics[width=0.45\textwidth]{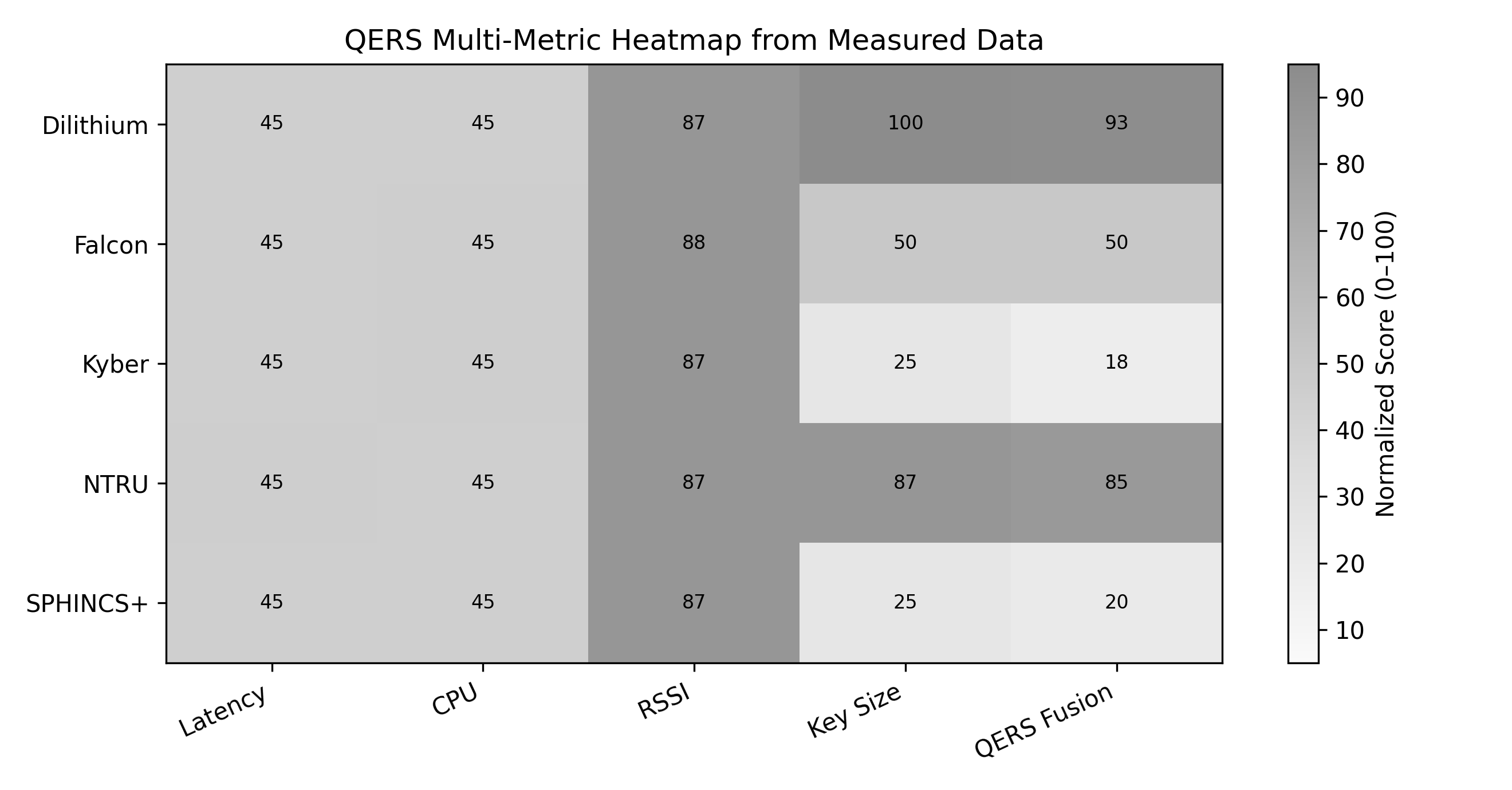}
    \caption{Scenario~1: QERS multi-metric heatmap (normalized 0–100).}
    \label{fig:heat1}
\end{figure}

\begin{figure}[htbp]
    \centering
    \includegraphics[width=0.45\textwidth]{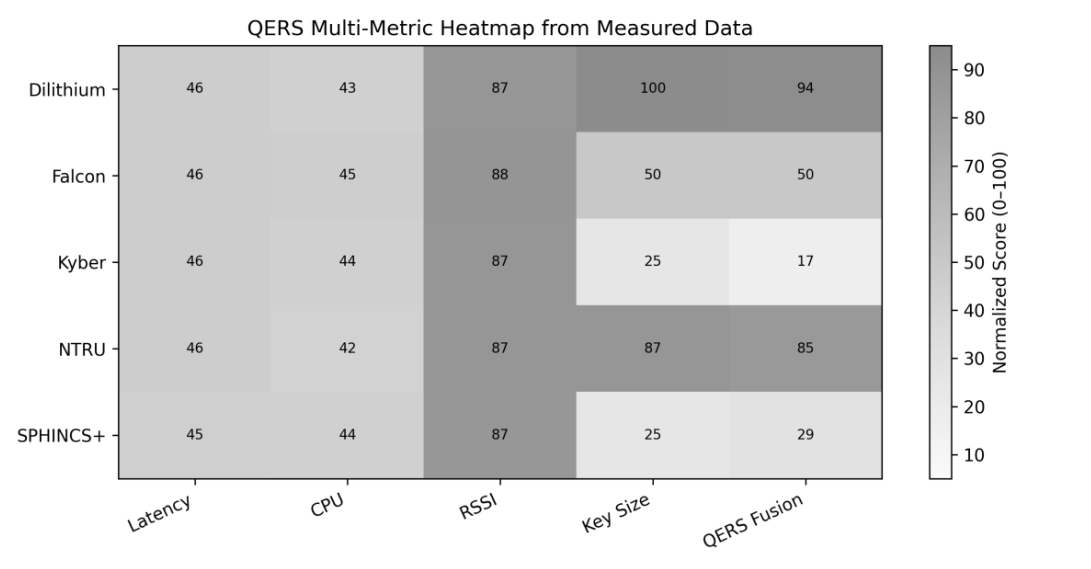}
    \caption{Scenario~2: QERS multi-metric heatmap (normalized 0–100).}
    \label{fig:heat2}
\end{figure}

Figures~\ref{fig:heat1} and~\ref{fig:heat2} visualize the normalized multi-metric inputs and resulting QERS behavior for both wireless scenarios. 
The heatmaps illustrate how QERS integrates latency, CPU usage, RSSI, and key-related effects into a composite resilience indicator. 
Variations across algorithms demonstrate the sensitivity of QERS to changes in cryptographic and wireless conditions.
\vspace{0.1 cm}

\vspace{0.1 cm}

\subsection{Distribution of Scores}
\begin{figure}[htbp]
    \centering
    \includegraphics[width=0.45\textwidth]{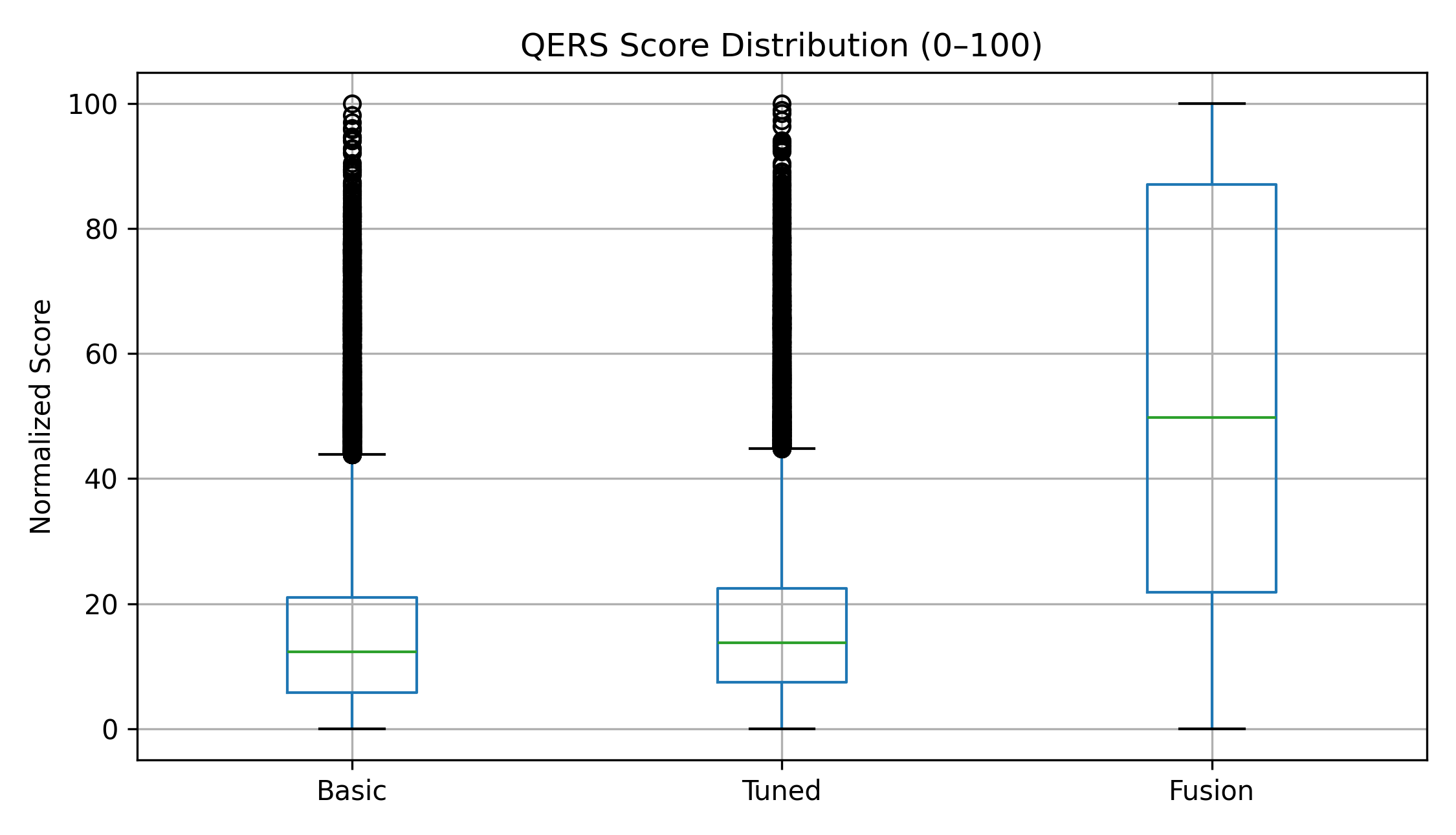}
    \caption{Scenario~1: Distribution of normalized QERS scores.}
    \label{fig:box1}
\end{figure}

\begin{figure}[htbp]
    \centering
    \includegraphics[width=0.45\textwidth]{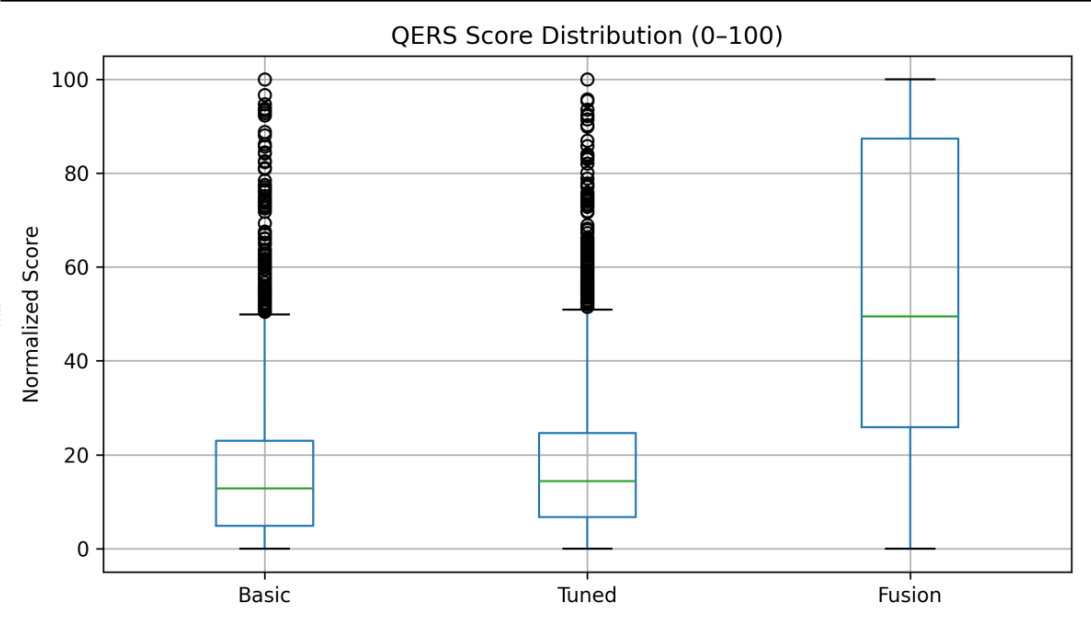}
    \caption{Scenario~2: Distribution of normalized QERS scores.}
    \label{fig:box2}
\end{figure}

Figures~\ref{fig:box1} and~\ref{fig:box2} show the distribution of QERS scores over the 24-hour period. 
The reduced spread of QERS$_{Fusion}$ illustrates the stabilizing effect of the fusion and smoothing layers when compared with raw or single-metric inputs.
\vspace{0.1 cm}

\subsection{Representative Scatter Analysis}
\begin{figure}[htbp]
    \centering
    \includegraphics[width=0.45\textwidth]{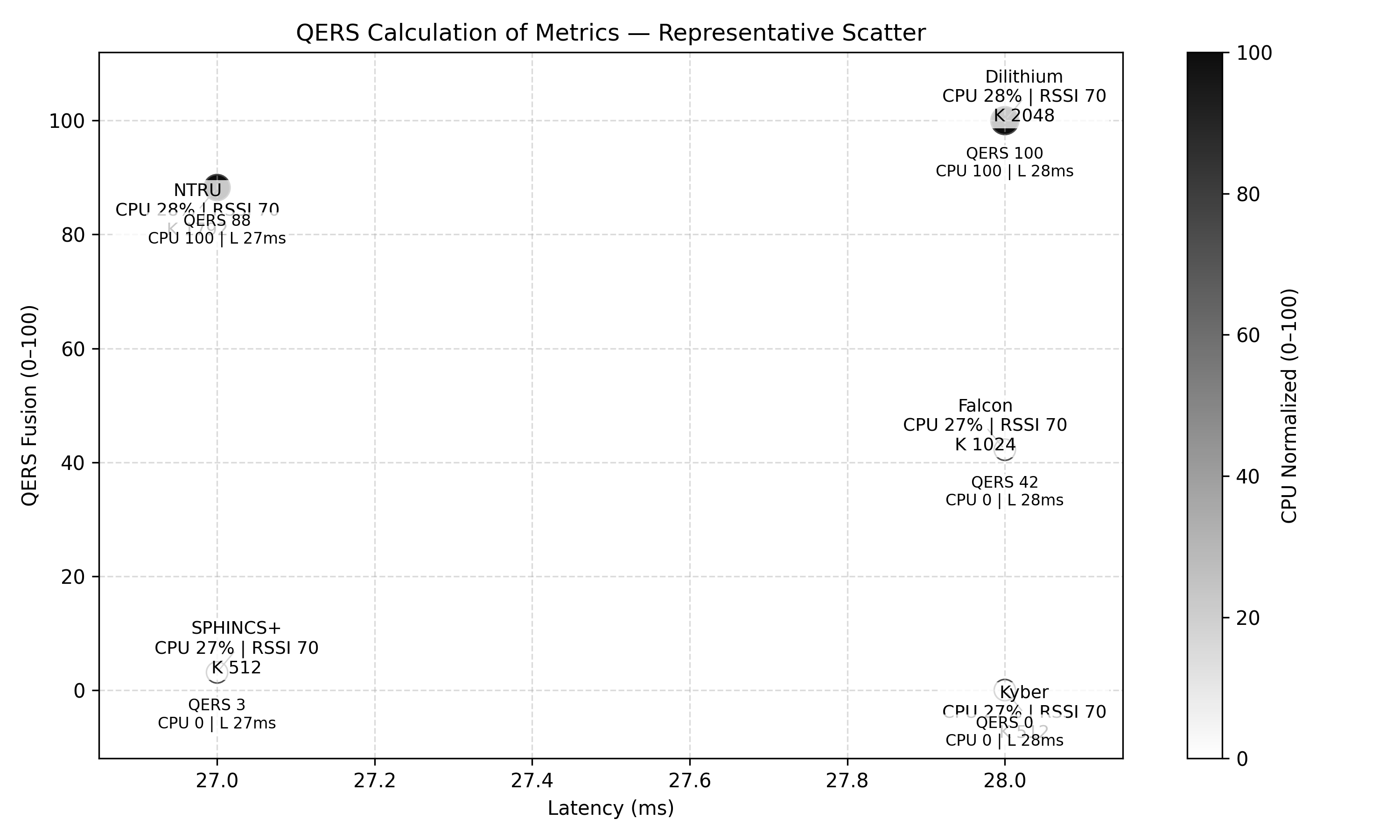}
    \caption{Scenario~1: Representative scatter (QERS vs latency).}
    \label{fig:scat1}
\end{figure}

\begin{figure}[htbp]
    \centering
    \includegraphics[width=0.45\textwidth]{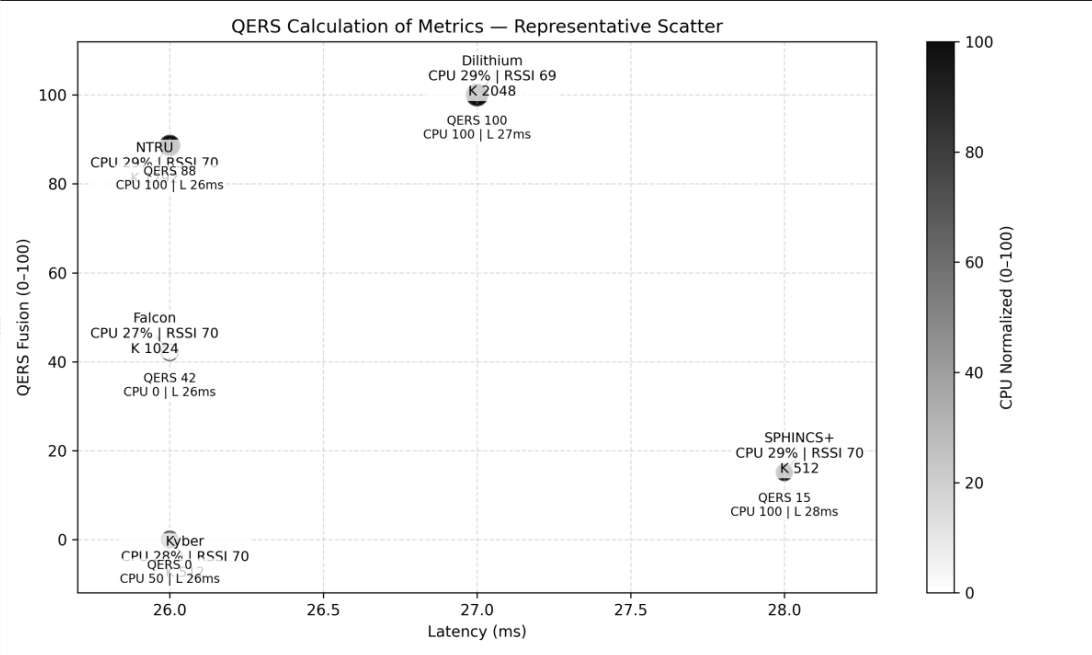}
    \caption{Scenario~2: Representative scatter (QERS vs latency).}
    \label{fig:scat2}
\end{figure}

Figures~\ref{fig:scat1} and~\ref{fig:scat2} provide representative scatter plots using median values per algorithm.  These plots demonstrate how QERS varies in relation to latency while preserving the relative structure of the underlying measurements.
\vspace{0.1 cm}

\vspace{0.1 cm}

\begin{table}[ht]
\centering
\caption{Average QERS$_{Fusion}$ scores for near and far communication scenarios.}
\label{tab:distance_summary}
\begin{tabular}{lcc}
\hline
\textbf{Algorithm} & \textbf{Close Range} & \textbf{5 m Distance} \\
\hline
Kyber      & 38.2 & 33.5 \\
Dilithium  & 72.4 & 69.8 \\
Falcon     & 55.1 & 50.7 \\
SPHINCS+   & 41.3 & 37.4 \\
NTRU       & 68.9 & 64.2 \\
\hline
\end{tabular}
\end{table}
\vspace{0.1 cm}

\subsection{Distance Comparison}
Table~\ref{tab:distance_summary} summarizes average QERS$_{Fusion}$ values across the two wireless scenarios. 
The reduction in QERS at increased distance illustrates that the proposed framework captures environmental stress in addition to cryptographic and computational effects.
\vspace{0.1 cm}

\section{Discussion}
The experimental results demonstrate that PQC schemes behave differently when evaluated in a resource-constrained IoT environment using a multi-metric scoring model such as QERS \cite{leelavathi2025smcrp}. Unlike traditional benchmarks that isolate individual parameters, QERS reveals how latency, CPU load, signal strength, and key size interact in practice \cite{silva2016case, bouramdane2023cyberattacks}.
\vspace{0.1 cm}

Across both scenarios, Dilithium and NTRU achieved the highest QERS$_{Fusion}$ scores. These algorithms benefit from strong security properties and predictable execution behavior, which results in stable normalized scores even when wireless conditions vary \cite{abbasi2025practical}. 

In contrast, Kyber and SPHINCS+ produced consistently lower QERS values. This outcome does not imply weaker cryptographic strength; rather, it reflects their design emphasis on reducing overhead, which lowers their contribution across certain QERS dimensions such as key size and computational weight \cite{magyari2025optimizing, nguyen2024efficient}.
\vspace{0.1 cm}

The distance experiment provided additional insight. Increasing the ESP32’s separation from the access point caused measurable reductions in QERS scores, particularly for algorithms that are more sensitive to retransmissions and signal fluctuations \cite{sklavos2017wireless}. This indicates that QERS captures environmental stress in addition to purely computational factors, which is essential for IoT deployments operating under unstable connectivity \cite{ukil2011embedded, zlatev2006computational}.
\vspace{0.1 cm}

Importantly, the Fusion layer did not eliminate variation entirely. Instead, it smoothed transient spikes while preserving real system behavior. This supports the interpretation of QERS as a decision-support indicator rather than a fixed or absolute performance rating \cite{kumar2020time}.
\vspace{0.1 cm}

\section{Conclusion}
This work introduced QERS, a practical and dynamic evaluation framework for assessing post-quantum cryptography in constrained IoT environments. Unlike classical benchmarking approaches that treat performance metrics in isolation, QERS integrates latency, CPU load, signal quality, and key size into a unified, normalized resilience score. A real ESP32-based testbed was used to collect continuous measurements across multiple PQC algorithms and wireless distances, demonstrating that QERS exposes performance and reliability trade-offs that are difficult to observe using single-metric analysis.

Future work will extend QERS with additional real-world indicators such as packet loss, temperature, and latency variance, and will evaluate its behavior across diverse IoT platforms. Overall, QERS provides a reproducible, explainable, and hardware-aware methodology for selecting post-quantum cryptographic algorithms that balance security, efficiency, and operational stability in real IoT deployments.

\bibliographystyle{IEEEtran}
\bibliography{bibfil}

\end{document}